\documentclass[preprint]{iucr}              
\usepackage{amsmath}
\usepackage{url}

     \journalcode{S}          

\begin{document}                  

\renewcommand\ack[1]{}  



\title{Stroboscopic detection of nuclear resonance in an arbitrary scattering channel}


\cauthor[a]{L.}{De\'{a}k}{deak.laszlo@wigner.mta.hu}{}
\author[a]{L.}{Botty\'{a}n}
\author[b]{R.}{Callens}
\author[b]{R.}{Coussement}
\aufn{Deceased 9 July 2012}
\author[c]{M.}{Major}
\aufn{On leave from Wigner RCP, RMKI, P.O.B.~49, 1525~\city{Budapest}, \country{Hungary}}
\author[d]{S.}{Nasu}
\aufn{Deceased 16 April 2014}
\author[b]{I.}{Serdons}
\author[e]{H.}{Spiering}
\author[f]{Y.}{Yoda}

\aff[a]{Wigner RCP, RMKI, P.O.B.~49, 1525~\city{Budapest}, \country{Hungary}}
\aff[b]{Instituut voor Kern- en Stralingsfysica, K.U.Leuven, Celestijnenlaan~200D, 
B-3001~\city{Leuven}, \country{Belgium}}
\aff[c]{Institute for Materials Science, Technische Universit\"{a}t Darmstadt, 64287 \city{Darmstadt}, 
\country{Germany}}
\aff[d]{Grad.\ School of Eng.\ Sci., Osaka Univ., Toyonaka, \city{Osaka} 560-8531, \country{Japan}}
\aff[e]{Institut f\"{u}r Anorganische und Analytische Chemie, Johannes Gutenberg
Universit\"{a}t Mainz, Staudinger Weg 9, D-55099 \city{Mainz}, \country{Germany}}
\aff[f]{SPring-8 JASRI, 1-1-1 Kouto, Mikazuki-cho, Sayo-gun, \city{Hyogo} 679-5198, \country{Japan}}









\maketitle                        

\begin{synopsis}
The theory of heterodyne/stroboscopic detection of nuclear resonance
scattering is developed for various dynamical scattering channels. 
The grazing incidence case is discussed in detail and is experimentally
demonstrated on magnetic multilayers.
\end{synopsis}

\begin{abstract}
The theory of heterodyne/stroboscopic detection of nuclear resonance
scattering is developed, starting from the total scattering matrix as a 
product of the matrix of the reference sample and the sample under study.
This general approach holds for any dynamical scattering channel.
The forward channel, which is discussed in detail in the literature,
reveals the speciality that
electronic scattering causes only an energy independent diminution of the
intensity. For all other channels, complex resonance line shapes
in the heterodyne/stroboscopic spectra---as a result of interference of
electronic and nuclear scattering---is encountered.
The grazing incidence case is evaluated and described in detail.
Experimetal data of classical grazing incidence reflection and their 
stroboscopic detection on $\left[ ^{\mathrm{nat}}\mathrm{Fe}/^{57}\mathrm{Fe}%
\right] _{10}$ and antiferromagnetic $\left[ ^{57}\mathrm{Fe}/\mathrm{Cr}%
\right] _{20}$ multilayers are fitted simultaneously.
\end{abstract}
\section{Introduction}
Nuclear resonant scattering (NRS) of synchrotron radiation (SR) has become
an established method for the study of nuclear hyperfine interaction during
the last two decades \cite{Gerdau2000,Roehlsberger2004}. The spectrum is
conventionally recorded as the time response of the nuclear ensemble
following a short resonant synchrotron pulse, which simultaneously excites
all resonant transitions between hyperfine-split nuclear sublevels. The
observed beating frequencies are characteristic for the hyperfine fields in
the specimen. As an alternative to nuclear resonant forward scattering of SR
in time domain, a heterodyne detection scheme was suggested
\cite{Coussement1996,Labbe2000}. The two scatterers, viz. the one
under investigation and as a reference sample a single-line M\"{o}ssbauer 
absorber, are mounted on a M\"{o}ssbauer drive. The heterodyne spectrum is
the full time integral of the delayed counts, plotted as a function of the
Doppler velocity of the reference sample. An advantage of this experimental
setup is the similarity of the spectra to those in the conventional
energy-domain M\"{o}ssbauer spectroscopy \cite{Coussement1996,Labbe2000}.
Although conventional M\"{o}ssbauer spectroscopy delivers similar
information on hyperfine interactions, the special properties of SR like
high collimation, high degree of polarization and high brilliance increase
the number of possible applications of NRS of SR.
Furthermore, the heterodyne setup allows for dense bunch
modes of the synchrotron (with bunch separation time much shorter than the
nuclear lifetime), which are not suitable for time differential NRS
experiments.

Undistorted time integration of the nuclear response can only be performed
if the large non-resonant intensity contribution is extinguished.
Experimentally, this can be achieved by using radiation from a nuclear
monochromator \cite{Smirnov1997} or by applying a polarizer/analizer setup 
\cite{Labbe2000}. An alternative approach, namely, ``stroboscopic detection'',
is based on appropriate time gating \cite{Callens2002,Callens2003},
i.e., integration of the delayed time response in a periodic time window. The
period $t_p$ of the observation time window after the SR pulse is chosen so
that $1/t_p$ falls within the frequency range of the hyperfine interactions
in the investigated specimen. This leads to new type of periodic resonances
at certain Doppler velocities that are shifted from the M\"{o}ssbauer
resonances by $mh/t_p$, with $h$ being Planck's constant, and $m$ an integer
number indicating the stroboscopic order \cite{Callens2002,Callens2003}. The
period $t_p$ should be selected according to the hyperfine spectral range,
the synchrotron bunch period and the detector dead time 
\cite{Inge2004,Callens2003}.

So far the theory of stroboscopic detection scheme has only been developed 
and discussed in detail for forward scattering geometry. The several applications 
of NRS in surface and thin-film magnetism that make use of the grazing incidence geometry 
\cite{Roehlsberger2004,Chumakov1999,Roehlsberger1999,Deak1999a,Roehlsberger2003,Sladecek2002}
call for computer programs that easily allow to fit data obtained by stroboscopic
detection as well. 
In this geometry, the interferences of the SR plane waves, scattered from the surface and
interfaces of a stratified sample, provide information on the value,
direction and topology of the internal fields in the sample with
nanometer depth resolution. 

Recently, interesting experiments have been performed using stroboscopic detection in
the grazing 
incidence case \cite{Roehlsberger2010,Roehlsberger2012}, which demonstrates the 
potential of this method. 

Grazing-incident NRS of SR, often called Synchrotron 
M\"{o}ssbauer Reflectometry (SMR) \cite{Gerdau2000,Deak01,Deak02}, has been
established in both time and angular regime \cite{Chumakov1999,Deak02,Nagy1999}, 
as time differential (TD) and time integral (TI) SMR, respectively. In the forward scattering 
channel, the prompt electronic scattering homogeneously contributes to the stroboscopic 
spectrum and does not affect the spectral shape. For 
other scattering channels, including grazing incidence reflection,
the interference of the electronic and nuclear scattering provides further information. 
The stroboscopic SMR line shape may considerably differ from the forward M\"{o}ssbauer
spectrum, calling for a specialized computer code.

The dynamical theory of x-ray scattering gives a self-consistent description
of the radiation field in all scattering channels of the system of
scatterers, taking all orders of multiple scattering into account. Theories
that expand the coherent elastic scattering to the case of sharp nuclear
resonances \cite{Afanasev1963,Kagan1964,Hannon68,Hannon69,Hannon3} have been
applied to
various scattering geometries. The simplest cases are the
one-beam cases, such as forward and off-Bragg scattering, and the two-beam
cases, the Bragg-Laue scattering \cite{Hannon69,Sturhahn1994} and the grazing
incidence scattering \cite{Roehlsberger2003,Hannon69,Hannon85,Andreeva1,Deak96}. 
In the grazing incidence limit, an optical model was derived from the dynamical theory 
\cite{Hannon69,Hannon85}, which has been implemented in numerical calculations 
\cite{Roehlsberger2003}. The reflectivity formulae given by \citeasnoun{Deak96} and 
\citeasnoun{Deak01} are suitable for fast numerical calculations in order to 
actually fit the experimental data \cite{Spiering00} and, as has been shown 
\cite{Deak1999}, this optical method is equivalent to that of the other approaches 
in the literature \cite{Roehlsberger2003,Hannon85}.

The aim of the present paper is to develop the concept of the
heterodyne/stro\-bo\-scop\-ic 
detection and to
establish the formula that can be applied to any scattering channel, 
like forward scattering, Bragg, off-Bragg and grazing incidence scattering.

This paper is
organized as follows. In the second section, the
heterodyne/stroboscopic intensity formula for the propagation of $\gamma $%
--photons in a medium containing both electronic and resonant nuclear
scatterers is derived. The equivalence to the previously discussed calculations for
the forward channel \cite{Callens2002,Callens2003}
are shown and 
the important specific case of the stroboscopic grazing incidence reflection
are outlined. In the third section, features of
the grazing incidence case
are demonstrated by least-squares fitted
experimental stroboscopic SMR spectra on isotope-periodic $\left[ ^{\mathrm{%
nat}}\mathrm{Fe}/^{57}\mathrm{Fe}\right] $ and antiferromagnetically ordered 
$\left[ ^{57}\mathrm{Fe}/\mathrm{Cr}\right] $ multilayer films.

\section{Heterodyne/Stroboscopic detection of Nuclear\newline
Resonance Scattering}

\subsection{General considerations}

The setup of a heterodyne/stroboscopic NRS of SR experiment includes two
scatterers, the investigated specimen and an additional reference sample
\cite{Coussement1996,Labbe2000,Callens2002,Callens2003}, the latter being
mounted on a M\"{o}ssbauer drive (in forward scattering geometry). The 
M\"{o}ssbauer drive provides a Doppler-shift $E_v=\left( v/c\right) E_0$ of the
nuclear energy levels, with $c,$ $v$ and $E_0$ being the velocity of light, 
the velocity of the drive and the energy of the M\"{o}ssbauer
transition, respectively.
The polarization dependence of the  nuclear scatterers is described 
adopting the notation of 
\citeasnoun{Sturhahn1994}, by $2\times 2$ transmissivity and reflectivity matrices
commonly called \textit{scattering matrices}.
The scattering of the synchrotron photons on
the specimen and the reference sample is described by the 
\textit{total scattering matrix} $T_\tau \left( E,E_v\right)$,
\begin{equation}
T_\tau \left( E,E_v\right) =T_\tau ^{\left( \mathrm{s}\right) }\left(
E\right) T^{\left( \mathrm{r}\right) }\left( E-E_v\right) ,  \label{tottrans}
\end{equation}
a product of the scattering matrices of the reference sample $T^{\left(
\mathrm{r}\right)} $ and of the 
investigated specimen $T^{\left(\mathrm{s}\right)}$ \cite{Blume68} in the energy domain.
The index $\tau$ specifies the open scattering channel \cite{Hannon69,Sturhahn1994}.
The scattering matrix 
$T^{\left( \mathrm{r}\right) }\left( E-E_v\right) \,$ of the reference sample
depends on the Doppler-shifted energy $E-E_v$, where the channel index $\tau $
is omitted for forward scattering. Note that both the electrons and 
the resonant M\"{o}ssbauer nuclei scatter the $\gamma $--photons coherently so
the scattering matrices have a resonant nuclear
and nearly energy-independent electronic contribution. At
energies being far from the M\"{o}ssbauer resonances $\left( E\rightarrow
\pm \infty \right) $ on a hyperfine scale, the individual scattering
matrices $T^{\left( \mathrm{s},\mathrm{r}\right) }\left( E\rightarrow \pm
\infty \right) $, and thus their product $T_\tau \left( E\rightarrow \infty
,E_v\right) \equiv T_{\tau ,\infty }\,$in Eq.~(\ref{tottrans}), approach
the non-resonant electronic contribution: 
\begin{equation}
T_{\tau ,\infty }=T_{\tau ,\mathrm{el}}^{\left( \mathrm{s}\right) }T_{%
\mathrm{el}}^{\left( \mathrm{r}\right) }.  \label{eltottrans}
\end{equation}
Since the reference is mounted in forward geometry, its scattering matrix $%
T^{\left( \mathrm{r}\right) }\left( E\right) $ is the matrix exponential
\begin{equation}
T^{\left( \mathrm{r}\right) }\left( E\right) =\exp \left[ ikd^{\left( 
\mathrm{r}\right) }n^{\left( \mathrm{r}\right) }\left( E\right) \right] ,
\label{Blumeexp}
\end{equation}
where $n^{\left( \mathrm{r}\right) }\left(E\right)$ is the index of refraction,
$d^{\left( \mathrm{r}\right) }$ the thickness and $k$ the vacuum
wave number of the incident radiation \cite{Blume68,Lax51}. The index of refraction 
is related to the susceptibility matrix $\chi$ \cite{Deak01,Deak96} through  
\begin{equation}
n^{\left( \mathrm{r}\right) }\left( E\right) \equiv I+\frac{\chi ^{\left( 
\mathrm{r}\right) }\left( E\right) }2,  \label{khidef}
\end{equation}
where $I$\ is the unit matrix and $\chi ^{\left( \mathrm{r}\right) }=\frac{4\pi
N^{\left( \mathrm{r}\right) }}{k^2}f^{\left( \mathrm{r}\right) }$, with $N^{\left( \mathrm{r}\right) }$
and $f^{\left( \mathrm{r}\right) }$ being the density of the scattering centers
and the $2\times 2$\ coherent forward scattering amplitude \cite{Blume68},
respectively. The susceptibility is the sum of the electronic and the
nuclear susceptibilities,
\begin{equation}
\chi ^{\left( \mathrm{r}\right) }\left( E\right) =\chi _{\mathrm{el}%
}^{\left( \mathrm{r}\right) }+\chi _{\mathrm{nuc}}^{\left( \mathrm{r}\right)
}\left( E\right) .  \label{khisum}
\end{equation}
With Eqs.~(\ref{Blumeexp}), (\ref{khidef}) and (\ref{khisum}), the
transmissivity of the reference is expressed as a product of electronic
and nuclear transmissivities,
\begin{equation}
T^{\left( \mathrm{r}\right) }\left( E\right) =T_{\mathrm{el}}^{\left( 
\mathrm{r}\right) }\tilde{T}_{\mathrm{nuc\vphantom{l}}}^{\left( \mathrm{r}\right)
}\left( E\right) ,  \label{sepel}
\end{equation}
where 
\begin{equation}
\tilde{T}_{\mathrm{nuc}}^{\left( \mathrm{r}\right) }\left( E\right) =\exp
\left( ikd^{\left( \mathrm{r}\right) }\frac{\chi _{\mathrm{nuc}}^{\left( 
\mathrm{r}\right) }\left( E\right) }2\right) .  \label{expnukl}
\end{equation}

$T_\tau^{\left(\mathrm{s}\right)}\left(E\right)$ is determined from the 
respective theory of wave propagation of channel $\tau $ (forward, Bragg-Laue,
grazing incidence, etc. scattering), i.e., from the dynamical theory
\cite{Hannon68,Hannon69,Hannon3,Sturhahn1994,Hannon85}.

In forward scattering, due to the
exponential expression in (\ref{Blumeexp}), the total transmissivity $
T\left( E,E_v\right) =T_\infty \tilde{T}_{\mathrm{nuc}}^{\left( \mathrm{r}
\right) }\left( E-E_v\right) \tilde{T}_{\mathrm{nuc}}^{\left( \mathrm{s}
\right) }\left( E\right) $ is proportional to $T_\infty$. Therefore, in
this special case, the electronic scattering is a simple multiplicative
factor, which does not affect the spectral shape. 

The intensity $I_{\tau }$ allowing for a general
polarization state of the incident beam, the $2\times 2$ polarization 
density matrix $\rho$ (\possessivecite{Blume68}), is given by 
\begin{equation}
I_{\tau }\left( E,E_{v}\right) =\mathrm{Tr}\left[ T_{\tau }^{\dagger }\left(
E,E_{v}\right) T_{\tau }\left( E,E_{v}\right) \rho \right] .
\label{Blumeint}
\end{equation}

The beating time response to a single short polychromatic photon bunch of SR  
is obtained by the Fourier transform of the energy domain scattering matrices
\cite{Gerdau2000,Hannon3},
\begin{equation}
T_{\tau }\left( t,E_{v}\right) =\frac{1}{\sqrt{2\pi }\hbar }\int \mathrm{d}E%
\text{\thinspace }\left[ T_{\tau }\left( E,E_{v}\right) -T_{\tau ,\infty }%
\right] \exp \left( -i\frac{E}{\hbar }t\right),   \label{timetrans}
\end{equation}
where, by subtracting the constant $T_{\tau ,\infty }$, the Dirac
delta--like prompt $(t=0)$ and the delayed $\left( t>0\right) $ time
responses are separated \cite{Sturhahn1994}. We note that Eq.~(\ref{timetrans})
is valid only for delayed times $t>0$ after the SR bunch $(t=0)$, but
$T_{\tau }\left( t,E_{v}\right) = 0$ for $t<0$\,!
In the same way as for Eq.~(\ref{Blumeint}),
the delayed intensity in time domain becomes
\begin{equation}
I_{\tau }\left( t,E_{v}\right) =\mathrm{Tr}\left[ T_{\tau }^{\dagger }\left(
t,E_{v}\right) T_{\tau }\left( t,E_{v}\right) \rho \right] .
\label{Blumeinttim}
\end{equation}%

For a heterodyne/stroboscopic NRS of SR experiment a time window function is introduced,
which can be described by boxcar functions, namely, $S(t)=1$ 
for $mt_{\mathrm{B}}+t_{1}<t<mt_{\mathrm{B}}+t_{2}$ and $S(t)=0$ otherwise, with a  time
interval $t_{\mathrm{B}}$ between the synchrotron bunches and an integer number $m$.
The periodic time window function is expanded in Fourier series,
\begin{equation}
S\left( t\right) =\sum_{-\infty }^{\infty }s_{m}\exp \left( im\Omega
t\right) ,  \label{fourwin}
\end{equation}%
where $\Omega =\tfrac{2\pi }{t_{\mathrm{B}}}$ is the angular frequency of
the SR bunches \cite{Callens2002,Callens2003}. 

The total delayed photon rate $D_{\tau }\left( E_{v}\right)$ of one bunch is 
\begin{equation}
D_{\tau }\left( E_{v}\right) =\int\limits_{-\infty }^{\infty }\mathrm{d}t%
\text{\thinspace }S\left( t\right) I_{\tau }(t,E_{v}),  \label{stroboint}
\end{equation}%
the integral of the intensity $I_{\tau }\left( t,E_{v}\right)$ times $S(t)$.
Since there is no coherence between photons generated by different electron 
bunches, the integral of the contribution of one bunch reveals the correct 
contribution of multiple bunches with periodicity of $t_{\mathrm{B}}$.

Combining Eqs.~(\ref{timetrans}), 
(\ref{Blumeinttim}), (\ref{stroboint}) and (\ref{fourwin}), the delayed count rate can be written as 
\begin{equation}
D_{\tau }\left( E_{v}\right) =\sum_{-\infty }^{\infty }s_{m}\delta _{\tau
,m}\left( E_{v}\right) ,  \label{finstrobo}
\end{equation}%
where 
\begin{equation}
\delta _{\tau ,m}\left( E_{v}\right) =\frac{1}{\hbar }\int \mathrm{d}E%
\mathrm{Tr}\left\{ \left[ T_{\tau }^{\dagger }\left( E-m\varepsilon
,E_{v}\right) -T_{\tau ,\infty }^{\dagger }\right] \left[ T_{\tau }\left(
E,E_{v}\right) -T_{\tau ,\infty }\right] \rho \right\}  \label{defalfa}
\end{equation}%
and 
\begin{equation}
\varepsilon =\hbar \Omega .  \label{epsdef}
\end{equation}%
Since the time window $S\left( t\right) $ and the intensity $D_{\tau }\left(
E_{v}\right) $ are real functions, $S_{m}^{}=S_{-m}^{\ast }$ and $\delta
_{m}^{}=\delta _{-m}^{\ast }$
hold, and Eq.~(\ref{finstrobo}) can
be rewritten as 
\begin{equation}
D_{\tau }\left( E_{v}\right) =s_{0}\delta _{\tau
,0}+\sum\limits_{m=1}^{\infty }2\mathrm{Re}\left( s_{m}\delta _{\tau
,m}\right) .  \label{riet2}
\end{equation}%
The result in Eqs.~(\ref{finstrobo})--(\ref{epsdef}) is a direct
generalization of the intensity formula (\ref{Blumeint}) to the
heterodyne/stroboscopic NRS of SR for any observed channel $\tau$ in the
applied experimental geometry. This expression has already been derived for
the case of forward scattering \cite{Callens2002,Callens2003}. The $m=0$ term
was called the ``heterodyne spectrum''
\cite{Coussement1996,Callens2002}, while the $m>0$ terms were called
``stroboscopic resonances'' 
of order $m$ \cite {Callens2002}. Nevertheless, the stroboscopic resonances are not restricted
to the forward scattering case. They also appear in other experimental
geometries, including, as we
 show below, in the grazing incidence
scattering geometry.

\subsection{Grazing incidence geometry}

In what follows, stroboscopic SMR spectra will be discussed. In terms of
the dynamical theory, grazing incidence is a two-beam case. The $\tau =0^{+}$
transmission and the $\tau =0^{-}$ reflection channels are open
\cite{Roehlsberger2004,Hannon69,Sturhahn1994}, the latter being observed in SMR.
Close to the electronic total reflection, the reflected intensity is high.
Therefore, SMR is an experimentally fairly instructive special case. The
reflection from the surface of the specimen is a multiple coherent
scattering process of the (SR) photons on atomic electrons and resonant 
M\"{o}ssbauer nuclei \cite{Deak01,Hannon85,Deak96}. Like in the forward
case, this scattering is independent of the atomic positions in the
reflecting medium, such that the scattering is described
by its index of refraction $n\left( E\right) $ \cite{Deak96,Lax51}. Henceforth, 
in compliance with the literature \cite{Roehlsberger2003,Deak01,Deak96}, 
in the general theory, the scattering matrix
$T_\tau ^{\left( \mathrm{s}\right) }\left( E\right) $
will be replaced by the $2\times 2$ reflectivity matrix $R^{\left( \mathrm{s}%
\right) }\left( E,\theta \right) $, where $\theta $ is the angle of 
incidence. This takes into account the interferences of the reflected
radiation from the surfaces and interfaces between the layers with different
refraction index. The methods of calculating the reflectivity matrix can be
found in the literature \cite{Roehlsberger2003,Deak01,Deak96}. Accordingly, 
the total scattering matrix of the specimen and the reference from 
Eq.~(\ref{tottrans}) is 
\begin{equation}
T\left( E,E_v,\theta \right) =R^{\left( \mathrm{s}\right) }\left( E,\theta
\right) T^{\left( \mathrm{r}\right) }\left( E-E_v\right) .
\label{reftottrans}
\end{equation}
Similarly, for energies being far from the M\"{o}ssbauer resonances, Eq.~(%
\ref{eltottrans}) reads 
\begin{equation}
T_\infty \left( \theta \right) =R_{\mathrm{el}}^{\left( \mathrm{s}\right)
}\left( \theta \right) T_{\mathrm{el}}^{\left( \mathrm{r}\right) }.
\label{refeltottrans}
\end{equation}
Inserting $T\left( E,E_v\right) $ and $T_\infty $ into Eq.~(\ref{finstrobo}%
), the delayed count rate $D\left( E_v,\theta \right)$ of the
heterodyne/stroboscopic spectrum for grazing incidence (stroboscopic SMR
intensity) on the specimen is calculated.

\begin{figure}
\includegraphics{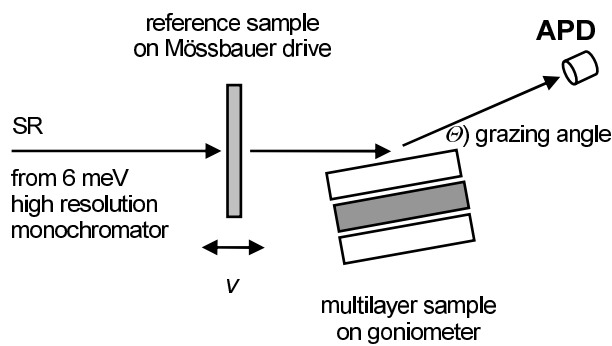}
\caption{Experimental setup for stroboscopic Synchrotron M\"{o}ssbauer Reflectometry.}
\label{expsetup}
\end{figure}

Combining Eqs.~(\ref{sepel}), (\ref{expnukl}) and (\ref{defalfa}) 
reveal
\begin{eqnarray}
\delta _m\left( E_v,\theta \right) & = & \frac{A^{\left( \mathrm{r}\right) }}%
\hbar \int \mathrm{d}E\mathrm{Tr}\left\{ \left[ \tilde{T}^{\dagger }\left(
E-E_v-m\varepsilon \right) R^{\dagger }\left( E-m\varepsilon \right) -R_{%
\mathrm{el}}^{\dagger }\right]\right.  \notag \\
& & \qquad \times \left.\left[ R\left( E\right) \tilde{T}\left( E-E_v\right) -R_{\mathrm{el}}%
\right] \rho \right\} ,  \label{dlform}
\end{eqnarray}
where $A^{\left( \mathrm{r}\right) }=\left| T_{\mathrm{el}}^{\left( \mathrm{r%
}\right) }\right| ^2$ is the electronic absorption of the reference sample.
For the sake of simplicity, the indices on the right hand side have been
omitted, so that $\tilde{T}_{\mathrm{nuc}}^{\left( \mathrm{r}\right)
}\rightarrow $$\tilde{T}$, $R_{\mathrm{el}}^{\left( \mathrm{s}\right)
}\left( \theta \right) \rightarrow R_{\mathrm{el}}$ and $R^{\left( \mathrm{s}%
\right) }\left( E,\theta \right) \rightarrow R\left( E\right) $. Note that
all reflectivities are those of the specimen, and all transmissivities are
those of the reference sample. With the relevant angular parameter $\theta $
for grazing incidence, Eq.~(\ref{finstrobo}) reads 
\begin{equation}
D\left( E_v,\theta \right) =\sum_{-\infty }^\infty s_m\delta _m\left(
E_v,\theta \right) .  \label{simpstrobo}
\end{equation}
The observed nuclear, as well as stroboscopic, resonances can be interpreted
in a straightforward manner using Eq.~(\ref{dlform}). Indeed, far from the
resonances, $R\left( E\rightarrow \infty \right) =R_{\mathrm{el}}$ and $%
\tilde{T}\left( E\rightarrow \infty \right) =1$, and the differences in the
square brackets in (\ref{dlform}) vanish. We expect a significant
contribution to the energy integral only if at least one energy argument of each
bracket is close to resonance, i.e., either 
\begin{subequations}
\label{co1}
\begin{align}
E-E_v-m\varepsilon &\simeq 0 \quad \text{and}  \label{co1a} \\
E &\simeq E_i  \label{co1b}
\end{align}
or 
\end{subequations}
\begin{subequations}
\label{co2}
\begin{align}
E-m\varepsilon & \simeq E_i \quad \text{and}  \label{co2a} \\
E-E_v & \simeq 0  \label{co2b}
\end{align}
are fulfilled, where $E_i$ is the energy of the
\textit{i}$^{\mathrm{th}}$ M\"{o}ssbauer 
resonance of the specimen. The $m^{\mathrm{th}}$ term of the sum in
Eq.~(\ref{simpstrobo}) contributes considerably if the Doppler velocity is
near to the corresponding shifted M\"{o}ssbauer resonance. In this case: 
\end{subequations}
\begin{subequations}
\label{fincond}
\begin{align}
E_v =E_i-m\varepsilon +\Delta ,  \label{finconda} \\
E_v =E_i+m\varepsilon +\Delta .  \label{fincondb}
\end{align}
Here, $\Delta $ is a small deviation (of the order of the resonance line width)\
from the energy $E_i-m\varepsilon $ or $E_i+m\varepsilon $, ensuring the
appearance of stroboscopic resonances also in grazing incidence geometry. In
the case of $m=0$, all four conditions of Eqs.~(\ref{co1}) and (\ref{co2})
may be true simultaneously. This means that, for $m=0$, nuclear scattering in
both samples, i.e., ``the radiative coupling of the samples'' 
\cite{Callens2003}, also contributes. Hence, the dynamical line broadening
(coherent speed-up) is the most effective in the heterodyne spectrum
(= baseline and resonances of stroboscopic order 0).

In order to perform computer simulations of stroboscopic spectra, Eqs.~(\ref%
{defalfa}) and (\ref{dlform}) were calculated for the forward scattering and
SMR cases, respectively. Eqs.~(\ref{defalfa}), (\ref{dlform}) and (\ref%
{riet2}) were implemented in the evaluation program EFFI \cite{Deak01,Spiering00}.
This program allows for least-square fitting
of stroboscopic spectra. Moreover, they can be fitted \textit{simultaneously}
with other types of spectra of the same specimen, such as forward
scattering, grazing incidence, conventional M\"{o}ssbauer and other spectra
of the implemented theory \cite{Deak01,Spiering00}. This way, the fit
constraints on the common parameters become very general, as already
described \cite{Deak01,Spiering00}.

\section{Experimental results and discussion}

In order to test the feasibility of this new reflectometric scheme, we investigated two film
specimens, a $^{\mathrm{nat}}\mathrm{Fe}/^{57}\mathrm{Fe}$ isotopic and a $%
^{57}$Fe/Cr antiferromagnetic multilayer, in grazing
incidence reflection geometry, using the $14.4~\mathrm{keV}$ M\"{o}ssbauer
transition of $^{57}$Fe nuclei. The experiments were performed at the BL09XU
nuclear resonance beam line of SPring-8 \cite{Yoda2001}. The
experimental setup is shown in Fig.~\ref{expsetup}. The synchrotron was
operated in the 203-bunch mode, corresponding to a bunch separation time of $%
23.6\ \mathrm{ns}$. The SR was monochromated by a Si(422)/Si(12~2~2) double
channel-cut high resolution monochromator with $6~\mathrm{meV}$ resolution.
It was incident on the K$_4$[$^{57}$Fe(CN)$_6$] single line pelleted
reference sample of effective thickness 11, and on the multilayer specimen
downstream mounted in grazing incidence geometry (Fig.~\ref{expsetup}). The 
M\"{o}ssbauer drive was operated in constant acceleration mode, with a
maximum velocity of $v_{\mathrm{max}}=20.24~\mathrm{mm/s}$. This maximum was
calibrated by fitting the velocity separation of the stroboscopic orders in
a forward scattering stroboscopic spectrum of a single line 
$^{57}\mathrm{Fe}$-enriched stainless steel absorber 
\cite{Callens2002,Callens2003}. The delayed radiation was detected using 
three $2~\mathrm{ns}$ dead time
Hamamatsu avalanche photo diodes (APD) in series. To record the delayed
intensity, a two-dimensional data acquisition system was used. Each count
was indexed according to the time elapsed after the synchrotron pulse (1024
channels), as well as to the velocity of the reference (1024 channels).
These stroboscopic SMR data were time integrated using appropriate time
windows of $t_p=7.87~\mathrm{ns}$ period and $3.93~\mathrm{ns}$ length
\cite{Callens2002,Callens2003}. Since the energy is measured in mm/s, the shift of
the first stroboscopic order, Eq.~(\ref{epsdef}), can be rewritten as 
\end{subequations}
\begin{equation}
\varepsilon ~\left[ \mathrm{mm}/\mathrm{s}\right] =1000\frac{\lambda ~\left[ 
\mathrm{nm}\right] }{t_p~\left[ \mathrm{ns}\right] }.  \label{stroboshift}
\end{equation}
With the wavelength $\lambda \approx 0.086~\mathrm{nm}$ for the M\"{o}%
ssbauer transition of $^{57}$Fe, the separation between the neighbouring
stroboscopic orders can be calculated to be $\varepsilon \approx 10.93~%
\mathrm{mm}/\mathrm{s}$. Note that this is the range of the hyperfine
splitting in case of $\alpha -\mathrm{Fe}$ (outer line separation is $10.62~%
\mathrm{mm}/\mathrm{s}$ at room temperature), and the stroboscopic orders
would only slightly overlap in case of a sample of low effective thickness
in forward scattering. However, in case of grazing incidence near the
critical angle of total external reflection due to the enhanced nuclear and
electronic multiple scattering, the M\"{o}ssbauer lines become extremely
broad and\ a strong overlap of the stroboscopic orders is expected. This
interference and partial overlap are manifested in rather complex resonance
line shapes and an intriguing angular dependence of the delayed intensity in
the various stroboscopic orders.

Both multilayers were prepared under ultra-high vacuum conditions by
molecular beam epitaxy at the IMBL facility in IKS Leuven. The\thinspace $[^{%
\mathrm{nat}}\mathrm{Fe}/^{57}\mathrm{Fe]}_{10}$ was prepared at room
temperature onto a Zerodur glass substrate. The first layer and all other $%
^{57}\mathrm{Fe}$-layers were 95.5\% isotopically enriched, and were
grown from a Knudsen cell. The natural Fe layers, which have a $^{57}%
\mathrm{Fe}$-concentration of 2.17\%, were grown from an electron gun
source. The nominal layer thickness was $3.15~\mathrm{nm}$ throughout the
multilayer stack for both $^{\mathrm{nat}}\mathrm{Fe}$ and $^{57}\mathrm{Fe}$%
. Conversion electron M\"{o}ssbauer spectra showed a pure $\alpha -\mathrm{Fe%
}$ spectrum. This spectrum was compared to a transmission M\"{o}ssbauer
spectroscopy spectrum of a natural iron calibration specimen, which was
provided by Amersham. Both hyperfine magnetic fields were fitted to be
identical within the experimental error of 0.04\%, and no sign of any second
phase contamination was found.

Preparation and characterization of the MgO(001)/[$^{57}$Fe/Cr]$_{20}$
multilayer sample has been described earlier 
\cite{BottyanBSF1,Nagy02a,Tancziko2004}. The layering was verified as epitaxial
and periodic, with thicknesses of $2.6~\mathrm{nm}$ for the $^{57}$Fe layer,
and $1.3~\mathrm{nm}$ for the Cr layer. SQUID magnetometry showed dominantly
antiferromagnetic coupling between neighboring Fe layers. According to
previous studies on this multilayer \cite{BottyanBSF1,Nagy02a,Tancziko2004},
the magnetizations in Fe align to the [100] and [010] perpendicular easy
directions in remanence, respectively corresponding to the [110] and 
[$\overline{1}$10] directions of the MgO substrate. The layer magnetizations
were aligned antiparallel in the consecutive Fe layers by applying a
magnetic field (1.6~T) above the saturation value (0.96~T) in the Fe[010]
easy direction of magnetization, and then releasing the field to remanence.
This alignment is global, the antiferromagnetic domains were only different
in the layer sequence of the parallel/antiparallel orientations 
\cite{Nagy02a}.

\subsection{Stroboscopic SMR on a $^{\mathrm{nat}}$Fe/$^{57}$Fe multilayer}

\begin{figure}
\includegraphics{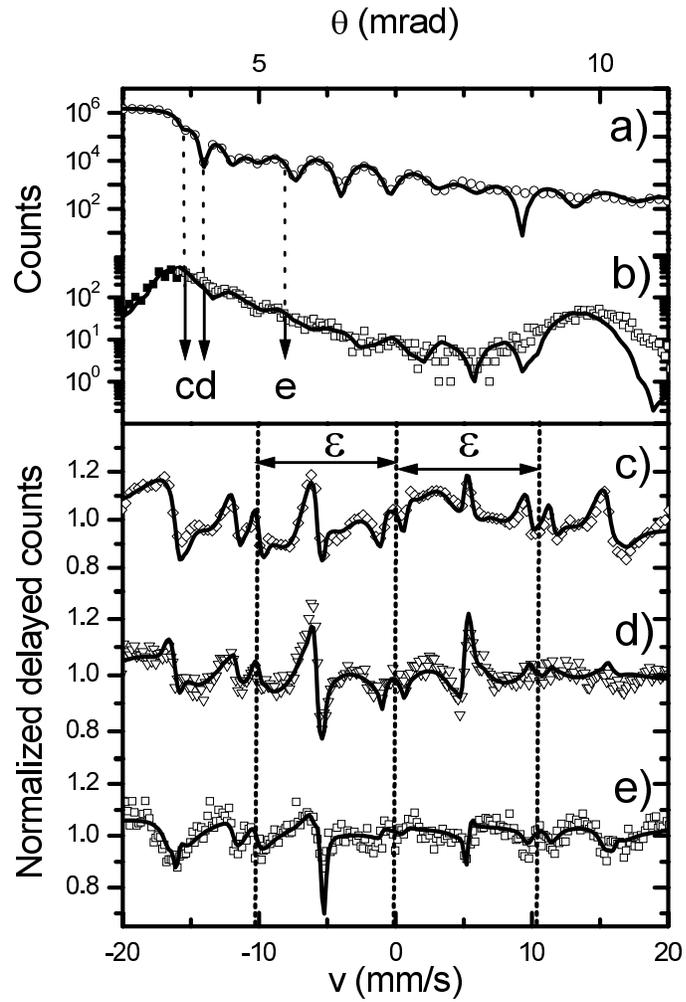}
\caption{Prompt electronic (a) and delayed nuclear reflectivity (b) curves
as well as stroboscopic SMR spectra (c) to e), of a $\left[ ^{\mathrm{nat}}%
\mathrm{Fe}/^{57}\mathrm{Fe}\right] _{10}$ isotopic multilayer at grazing
angles indicated by the arrows. Vertical dotted lines in panels c) to e)
indicate the center of the zero and $\pm 1$ order stroboscopic bands
separated by $\protect\varepsilon \approx 10.93~\mathrm{mm}/\mathrm{s}$ for
the applied observation window period.}
\label{fefe}
\end{figure}

Since in a $^{\mathrm{nat}}\mathrm{Fe}/^{57}\mathrm{Fe}$ isotope-periodic
multilayer the hyperfine field of $^{57}\mathrm{Fe}$ is that 
of $\alpha -\mathrm{Fe}$ throughout the sample, this multilayer is 
particularly suitable for studying the modification of the resonance 
line shapes due to interference between nuclear and electronic scattering 
\cite{Deak1999a,Deak1994,Chumakov1991,Chumakov1993}. Fig.~\ref{fefe} shows 
results for the multilayer saturated in a transversal magnetic field of 50 mT. 
Panel \textit{a} and \textit{b} give the prompt electronic and delayed TISMR curves,
respectively. The stroboscopic SMR spectra at the angles indicated by the
arrows are given in panel \textit{c} to \textit{e}. The peak in the delayed
reflectivity at the total reflection angle in panel \textit{b} is a special
feature of SMR described earlier \cite{Chumakov1999,Deak1994,Baron1994}. In
panels \textit{c} to \textit{e}, the four resonance lines of the $+1$ and $-1$
stroboscopic orders (right and left sides, respectively) partially overlap
with the $0^{\mathrm{th}}$ order in the central part of the spectrum.

The delicate interplay between electronic and nuclear scattering is
demonstrated by the considerable difference between the stroboscopic SMR
spectra \textit{c} to \textit{e} in Fig.~\ref{fefe}, which are taken at only
slightly different grazing angles. In contrast to the symmetric forward
scattering spectra \cite{Callens2002,Callens2003}, the stroboscopic SMR
spectra are asymmetric due to the interference between the electronic and
nuclear scattering. They also display both ``absorption-like'' and
``dispersion-like'' resonance line shape contributions. In case of decreased
nuclear scattering strength and of the same electronic reflectivity (cf.
panels \textit{d} and \textit{e} in Fig.~\ref{fefe}), the signal to baseline
ratio of the central part (heterodyne spectrum) decreases as compared to the
signal to baseline ratio of stroboscopic orders $\pm 1$\ in the spectrum
wings.

The full lines are simultaneous least squares fits, using the theory
outlined above and the computer code EFFI \cite{Spiering00}. The
interference between nuclear and electronic scattering makes it possible to
fit the layer structure in this isotope-periodic multilayer. The fitted
value of the total thickness of pure $\alpha -\mathrm{Fe}$ is $42.5\ \mathrm{%
nm,}$ comprised of nine times $1.49\ \mathrm{nm}$ of $^{\mathrm{nat}}\mathrm{%
Fe}$ and $3.23\ \mathrm{nm}$ of $^{57}\mathrm{Fe,}$ with $0.4\ \mathrm{nm}$
common roughness at the interfaces. In order to achieve the simultaneous
fit, displayed by the full line in Figure \ref{fefe}, we had to assume that
half a bilayer on top and bottom ($^{\mathrm{nat}}\mathrm{Fe}$ and $^{57}%
\mathrm{Fe}$, respectively) was modified. The transversal hyperfine magnetic
field was fixed to $33.08\ \mathrm{T}$ in the nine $^{57}\mathrm{Fe}$/$^{%
\mathrm{nat}}\mathrm{Fe}$ bi-layers in the middle of the multilayer, which
is the room temperature value for $\alpha -\mathrm{Fe}$.

\subsection{Stroboscopic SMR of an antiferromagnetic $^{57}$Fe/Cr multilayer}

\begin{figure}
\includegraphics{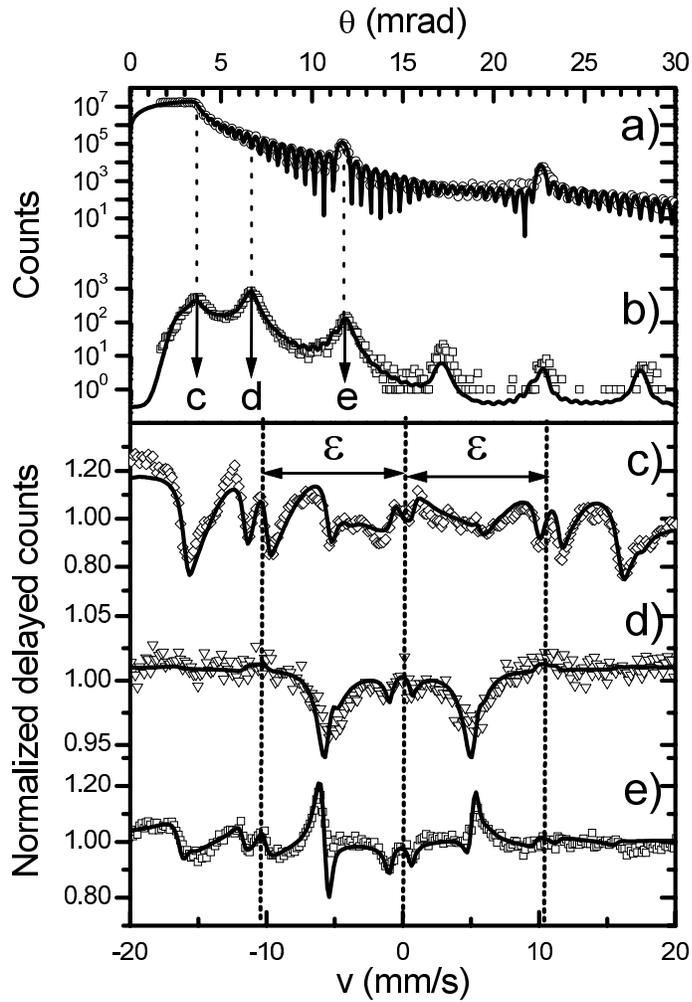}
\caption{Prompt electronic (a) and delayed nuclear (b) reflectivity curves
as well as stroboscopic SMR spectra (c to e) of a $\mathrm{MgO(001)}/\left[
^{57}\mathrm{Fe}/\mathrm{Cr}\right] _{20}$ antiferromagnetic multilayer at
various angles indicated by arrows in b). The consecutive Fe layer
magnetizations are aligned parallel/antiparallel with to the SR beam.
Vertical dotted lines in panels c) to e) indicate the center of the zero and 
$\pm 1$ order stroboscopic bands separated by $\protect\varepsilon \approx
10.93~\mathrm{mm}/\mathrm{s}$ for the applied observation window period.}
\label{fecraf}
\end{figure}

\begin{figure}
\includegraphics{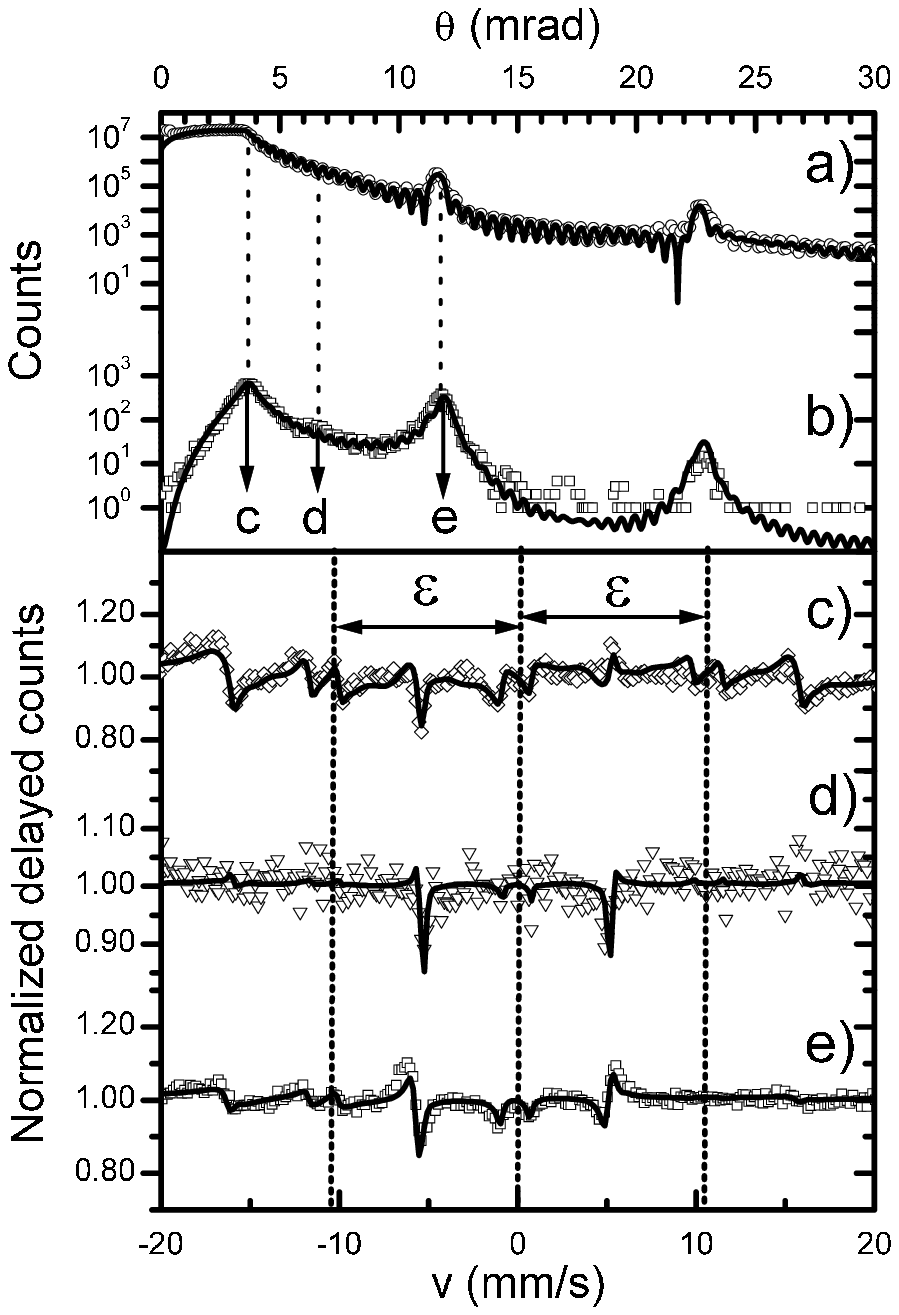}
\caption{Prompt electronic (a) and delayed nuclear (b) reflectivity curves
as well as stroboscopic SMR spectra (c to e) of a $\left[ ^{57}\mathrm{Fe}%
\left( 2.6\,\mathrm{nm}\right) /\mathrm{Cr}\left( 1.3\,\mathrm{nm}\right) %
\right] _{20}/\mathrm{MgO}$ antiferromagnetic multilayer at various angles
indicated by arrows. The consecutive Fe layer magnetizations are aligned
perpendicular to the SR beam. Vertical dotted lines in panels c) to e)
indicate the center of the zero and $\pm 1$ order stroboscopic bands
separated by $\protect\varepsilon \approx 10.93~\mathrm{mm}/\mathrm{s}$ for
the applied observation window period.}
\label{fecrfm}
\end{figure}

Fig.~\ref{fecraf} and Fig.~\ref{fecrfm} display similar sets of spectra of
an $^{57}$Fe/Cr antiferromagnetically coupled epitaxial multilayer on
MgO(001). The dots are the experimental data points, while the continuous
lines are simultaneous fits to a model structure of $\big[ {}^{57}\mathrm{Fe}%
\left( 2.6~\mathrm{nm}\right) /$ $\mathrm{Cr}\left( 1.3~\mathrm{nm}\right) %
\big] {}^{}_{20}\:\!$, based on the respective theory.

Non-resonant reflectivity, TISMR and stroboscopic SMR spectra were recorded
first with the Fe layer magnetizations parallel/antiparallel (Fig.~\ref%
{fecraf}) to the $k-$vector of the SR beam. The stroboscopic spectra were
taken at the angles of total reflection (c), at the antiferromagnetic (d)
and at the structural Bragg peak (e) positions. After this, a magnetic field
of 20 mT was applied to the multilayer in longitudinal direction. This is
known to flop the magnetizations to the perpendicular Fe(010) easy axis 
of the magnetization \cite{BottyanBSF1,Tancziko2004}. Non-resonant
reflectivity, TISMR and stroboscopic SMR spectra at the same angular
positions were again collected (Fig.~\ref{fecrfm}).

The major difference between Figs.~\ref{fecraf} and \ref{fecrfm} is the
presence, respectively absence, of the AF Bragg peak in the delayed
reflectivity curve b. This antiferromagnetic alignment, i.e., the
longitudinal hyperfine field of alternating sign in consecutive Fe layers,
is justified by the simultaneous fit in Fig.~\ref{fecraf}. In Fig.~\ref%
{fecrfm}, the fitted Fe magnetizations are perpendicular to the wave vector
of the SR. Indeed, the scattering amplitudes depend on the angle of the wave
vector and the direction of the hyperfine magnetic field. In the case of
perpendicular orientation, this angle is 90 degrees for consecutive layer
magnetizations and no AF contrast can be observed. In case of
parallel/anti--parallel orientations, however, the angles with respect to
the wave vector of SR are 0 and 180 degrees, respectively. Therefore, the
hyperfine contrast is present and the AF Bragg peak is visible in panel b of
Fig.~\ref{fecraf}.

The count rate at the baseline of a stroboscopic SMR spectrum, measured at a
certain grazing angle $\theta $, is closely related to the TISMR spectrum at
this angle. Therefore, the respective experimental count rates of the
stroboscopic SMR spectrum at the AF Bragg peak positions (panel d) differ by
almost two orders of magnitude. Spectrum 3d is also the only spectrum for
which no considerable enhanced dynamic broadening can be observed.

Note that, in panels d, the zeroth order resonances are considerably enhanced
with respect to the $\pm 1\,$order stroboscopic resonances. This can be
explained by an enhanced radiative coupling of the samples. Since the
radiative coupling does not contribute to the $\pm 1\,$order stroboscopic
resonances, it only influences the baseline and the central resonances.

At the multilayer Bragg reflections (panel e), and at the total reflection
peak (panel c), the suppression of the higher stroboscopic orders is much
smaller, which means that the radiative coupling term is not dominating here.
These spectra also show a left/right asymmetry due to the variation of the
phase of the total scattering amplitude with energy. This latter allows for
phase determination of the scattering amplitude from a set of stroboscopic
SMR spectra, which work will be published later.

\section{Summary}

In summary, the concept of heterodyne/stroboscopic detection of nuclear
resonance scattering was outlined for a general scattering channel, with
special emphasis on the grazing incidence reflection case. In any
non-forward scattering channel, the electronic scattering influences the NRS
spectral shape, while in forward scattering, this is a mere multiplicative
factor. The interplay between electronic and nuclear scattering, as a
function of the scattering angle, facilitates the determination of the
electronic and nuclear scattering amplitudes. The code of the present theory
has been merged into the EFFI program \cite{Spiering00}, and was used in
simultaneous data fitting of x-ray reflectivity, time integral reflectivity
and stroboscopic SMR spectra. Similar to time differential SMR, stroboscopic
SMR spectra have been shown to be sensitive to the direction of the
hyperfine fields of the individual layers. Therefore, it is possible to
apply this method to the study of magnetic multilayers and thin films. The
experiments on $\left[ ^{57}\mathrm{Fe}\left( 2.6~\mathrm{nm}\right) /%
\mathrm{Cr}\left( 1.3~\mathrm{nm}\right) \right] _{20}$ and $\left[ ^{%
\mathrm{nat}}\mathrm{Fe}/^{57}\mathrm{Fe}\right] _{10}$ multilayers
demonstrated that stroboscopic detection of synchrotron M\"{o}ssbauer
reflectometry of $^{57}$Fe-containing thin films is feasible in dense bunch
modes,
which are not necessarily suitable for time differential nuclear resonance
scattering experiments on $^{57}$Fe.

%
%
%
%
%
%
%
%


\ack
{The authors gratefully acknowledge the beam time supplied free of charge by
the Japan Synchrotron Radiation Institute (JASRI) for experiment No:
2002B239-ND3-np. Our gratitude goes to Dr. A.Q. Baron (SPring-8, JASRI) for
kindly supplying the fast Hammamatsu APD detectors and Dr. Johan Dekoster
(IKS\ Leuven) for preparing the multilayer samples for the experiment.
Support by the Flemish-Hungarian inter-governmental project BIL14/2002, the
DYNASYNC Framework Six project of the European Commission (contract number:
NMP4-CT-2003-001516) the Fund for Scientific research--Flanders (G.0224.02),
the Inter-University Attraction Pole (IUAP P5/1) and the Concerted Action of
the KULeuven (GOA/2004/02) is gratefully acknowledged. L.~De\'{a}k and R.~Callens 
thank the Deutscher Akademischer Austauschdienst (DAAD) and the
FWO-Flanders, respectively, for financial support.}


\referencelist[oldstrobo-140807]




\end{document}